\documentclass[12pt,a4paper]{article}

\usepackage{subfigure}         
\usepackage{comment}
\usepackage{float}              
\floatstyle{plaintop}
\restylefloat{table}

\usepackage{footmisc}
\usepackage{array}
\usepackage{footnote}
\usepackage[font={small}]{caption} 

\def\jpsidec{\jpsi \ra \mumu}
\def\phidec{\phi \ra \Kp \Km}
\def\Dsdec{\Dsm \ra \Kp \Km \pim}
\def\Bcdec{\Bc \ra \Bs \pip}

\def\Bcdecfulld{\Bc \ra \Bs(\ra \Dsm \pip) \pip}
\def\Bcdecfullj{\Bc \ra \Bs(\ra \jpsi \phi) \pip}
\def\Bsdecj{\Bs \ra \jpsi \phi}
\def\Bsdecd{\Bs \ra \Dsm\pip}

\def\frat{\frac{\sigma(\Bc)}{\sigma(\Bs)}}

\def\Bsstar{B_s^{*}}
\def\DsPi{D_s^-\pi^+}
\def\JpsiPhi{\jpsi\phi}

\restylefloat{table}
\usepackage{ifthen} 
\newboolean{pdflatex}
\setboolean{pdflatex}{true} 

\newboolean{articletitles}
\setboolean{articletitles}{true} 

\newboolean{uprightparticles}
\setboolean{uprightparticles}{false} 

\newboolean{inbibliography}
\setboolean{inbibliography}{false} 

\usepackage{microtype}
\usepackage{lineno}  
\usepackage{xspace} 

\usepackage{graphicx}  
\usepackage{color}
\usepackage{colortbl}
\graphicspath{{./figs/}} 

\usepackage{amsmath} 
\usepackage{amssymb}
\usepackage{amsfonts}
\usepackage{upgreek} 

\newcommand*\patchAmsMathEnvironmentForLineno[1]{%
\expandafter\let\csname old#1\expandafter\endcsname\csname #1\endcsname
\expandafter\let\csname oldend#1\expandafter\endcsname\csname
end#1\endcsname
 \renewenvironment{#1}%
   {\linenomath\csname old#1\endcsname}%
   {\csname oldend#1\endcsname\endlinenomath}%
}
\newcommand*\patchBothAmsMathEnvironmentsForLineno[1]{%
  \patchAmsMathEnvironmentForLineno{#1}%
  \patchAmsMathEnvironmentForLineno{#1*}%
}
\AtBeginDocument{%
\patchBothAmsMathEnvironmentsForLineno{equation}%
\patchBothAmsMathEnvironmentsForLineno{align}%
\patchBothAmsMathEnvironmentsForLineno{flalign}%
\patchBothAmsMathEnvironmentsForLineno{alignat}%
\patchBothAmsMathEnvironmentsForLineno{gather}%
\patchBothAmsMathEnvironmentsForLineno{multline}%
}

\usepackage{hyperref}    
\usepackage[all]{hypcap} 

\def\lhcb {\mbox{LHCb}\xspace}

\ifthenelse{\boolean{uprightparticles}}%
{

 \def\Pmu         {\ensuremath{\upmu}\xspace}

 \def\Ppi         {\ensuremath{\uppi}\xspace}

 \def\Ppsi        {\ensuremath{\uppsi}\xspace}

 \def\PDelta      {\ensuremath{\Delta}\xspace}                 
 \def\PXi      {\ensuremath{\Xi}\xspace}                 
 \def\PLambda      {\ensuremath{\Lambda}\xspace}                 
 \def\PSigma      {\ensuremath{\Sigma}\xspace}                 
 \def\POmega      {\ensuremath{\Omega}\xspace}                 
 \def\PUpsilon      {\ensuremath{\Upsilon}\xspace}                 
 
 \def\PB      {\ensuremath{\mathrm{B}}\xspace}                 
                  
 \def\PD      {\ensuremath{\mathrm{D}}\xspace}

 \def\PJ      {\ensuremath{\mathrm{J}}\xspace}                 
 \def\PK      {\ensuremath{\mathrm{K}}\xspace}

 \def\Pc      {\ensuremath{\mathrm{c}}\xspace}

 \def\Pi      {\ensuremath{\mathrm{i}}\xspace}

 \def\Ps      {\ensuremath{\mathrm{s}}\xspace}

}
{

 \def\Pmu         {\ensuremath{\mu}\xspace}

 \def\Ppi         {\ensuremath{\pi}\xspace}

 \def\Ppsi        {\ensuremath{\psi}\xspace}                 
                  
 \mathchardef\PDelta="7101
 \mathchardef\PXi="7104
 \mathchardef\PLambda="7103
 \mathchardef\PSigma="7106
 \mathchardef\POmega="710A
 \mathchardef\PUpsilon="7107
                  
 \def\PB      {\ensuremath{B}\xspace}                 
                  
 \def\PD      {\ensuremath{D}\xspace}

 \def\PJ      {\ensuremath{J}\xspace}                 
 \def\PK      {\ensuremath{K}\xspace}

 \def\Pc      {\ensuremath{c}\xspace}

 \def\Pi      {\ensuremath{i}\xspace}

 \def\Ps      {\ensuremath{s}\xspace}

}

\def\mumu       {\ensuremath{\Pmu^+\Pmu^-}\xspace}

\def\squark    {\ensuremath{\Ps}\xspace}

\def\cquark    {\ensuremath{\Pc}\xspace}

\def\pion  {\ensuremath{\Ppi}\xspace}

\def\pip   {\ensuremath{\pion^+}\xspace}
\def\pim   {\ensuremath{\pion^-}\xspace}

\def\kaon  {\ensuremath{\PK}\xspace}
  \def\Kbar  {\kern 0.2em\overline{\kern -0.2em \PK}{}\xspace}

\def\Kp    {\ensuremath{\kaon^+}\xspace}
\def\Km    {\ensuremath{\kaon^-}\xspace}

  \def\Dbar    {\kern 0.2em\overline{\kern -0.2em \PD}{}\xspace}
\def\D       {\ensuremath{\PD}\xspace}

\def\Dm      {\ensuremath{\D^-}\xspace}

\def\Dsm     {\ensuremath{\D^-_\squark}\xspace}

\def\Dssm    {\ensuremath{\D^{*-}_\squark}\xspace}

\def\B       {\ensuremath{\PB}\xspace}
\def\Bbar    {\ensuremath{\kern 0.18em\overline{\kern -0.18em \PB}{}}\xspace}

\def\Bz      {\ensuremath{\B^0}\xspace}

\def\Bu      {\ensuremath{\B^+}\xspace}

\def\Bs      {\ensuremath{\B^0_\squark}\xspace}

\def\Bc      {\ensuremath{\B_\cquark^+}\xspace}

\def\jpsi     {\ensuremath{{\PJ\mskip -3mu/\mskip -2mu\Ppsi\mskip 2mu}}\xspace}

  \def\Y#1S{\ensuremath{\PUpsilon{(#1S)}}\xspace}

\def\Lbar {\ensuremath{\kern 0.1em\overline{\kern -0.1em\PLambda}}\xspace}

\def\BF         {{\ensuremath{\cal B}\xspace}}

\def\BR         {\BF}
\newcommand{\decay}[2]{\ensuremath{#1\!\to #2}\xspace}         
\def\ra                 {\ensuremath{\rightarrow}\xspace}
\def\to                 {\ensuremath{\rightarrow}\xspace}

\def\BsToJPsiPhi  {\decay{\Bs}{\jpsi\phi}}

\def\AT#1     {\ensuremath{A_{\mathrm{T}}^{#1}}\xspace}           

\def\C#1      {\ensuremath{\mathcal{C}_{#1}}\xspace}                       
\def\Cp#1     {\ensuremath{\mathcal{C}_{#1}^{'}}\xspace}                    
\def\Ceff#1   {\ensuremath{\mathcal{C}_{#1}^{\mathrm{(eff)}}}\xspace}        
\def\Cpeff#1  {\ensuremath{\mathcal{C}_{#1}^{'\mathrm{(eff)}}}\xspace}       
\def\Ope#1    {\ensuremath{\mathcal{O}_{#1}}\xspace}                       
\def\Opep#1   {\ensuremath{\mathcal{O}_{#1}^{'}}\xspace}                    

\newcommand{\tev}{\ifthenelse{\boolean{inbibliography}}{\ensuremath{~T\kern -0.05em eV}\xspace}{\ensuremath{\mathrm{\,Te\kern -0.1em V}}\xspace}}
\newcommand{\gev}{\ensuremath{\mathrm{\,Ge\kern -0.1em V}}\xspace}
\newcommand{\mev}{\ensuremath{\mathrm{\,Me\kern -0.1em V}}\xspace}
\newcommand{\kev}{\ensuremath{\mathrm{\,ke\kern -0.1em V}}\xspace}
\newcommand{\ev}{\ensuremath{\mathrm{\,e\kern -0.1em V}}\xspace}
\newcommand{\gevc}{\ensuremath{{\mathrm{\,Ge\kern -0.1em V\!/}c}}\xspace}
\newcommand{\mevc}{\ensuremath{{\mathrm{\,Me\kern -0.1em V\!/}c}}\xspace}
\newcommand{\gevcc}{\ensuremath{{\mathrm{\,Ge\kern -0.1em V\!/}c^2}}\xspace}
\newcommand{\gevgevcccc}{\ensuremath{{\mathrm{\,Ge\kern -0.1em V^2\!/}c^4}}\xspace}
\newcommand{\mevcc}{\ensuremath{{\mathrm{\,Me\kern -0.1em V\!/}c^2}}\xspace}

\def\mum  {\ensuremath{\,\upmu\rm m}\xspace}

\def\invfb   {\ensuremath{\mbox{\,fb}^{-1}}\xspace}

\newcommand{\chisq}{\ensuremath{\chi^2}\xspace}

\def\gsim{{~\raise.15em\hbox{$>$}\kern-.85em
          \lower.35em\hbox{$\sim$}~}\xspace}
\def\lsim{{~\raise.15em\hbox{$<$}\kern-.85em
          \lower.35em\hbox{$\sim$}~}\xspace}

\def\pt         {\mbox{$p_{\rm T}$}\xspace}
\def\et         {\mbox{$E_{\rm T}$}\xspace}

\def\bcvegpy    {\mbox{\textsc{Bcvegpy}}\xspace}

\def\evtgen     {\mbox{\textsc{EvtGen}}\xspace}

\def\geant      {\mbox{\textsc{Geant4}}\xspace}

\def\photos     {\mbox{\textsc{Photos}}\xspace}

\def\pythia     {\mbox{\textsc{Pythia}}\xspace}

\def\tell1  {TELL1\xspace}
\def\ukl1   {UKL1\xspace}

\newcommand{\eg}{\mbox{\itshape e.g.}\xspace}
\newcommand{\ie}{\mbox{\itshape i.e.}\xspace}

\usepackage{cite} 
\usepackage{mciteplus}

\usepackage{longtable} 

\newcommand{\BcBsPiJpsiPhiResult}{
  \ensuremath{\left(2.20 \pm 0.49 \,(\text{stat})
      \pm 0.23 \,(\text{syst})\right) \times 10^{-3}}
}

\newcommand{\BcBsPiDsPiResult}{
  \ensuremath{\left(2.54 \pm 0.40 \,(\text{stat}) 
       ^{+0.23}_{-0.17} \,(\text{syst})\right) \times 10^{-3}}
}

\newcommand{\FinalResult}{
  \ensuremath{\left(2.37  \pm 0.31 \,(\text{stat})\,  
       \pm 0.11 \,(\text{syst})\,
       ^{+0.17}_{-0.13} \,(\tau_{\Bc})\right) \times 10^{-3}}
}

\newcommand{\FinalResultSevenTeV}{
  \ensuremath{\left(1.27  \pm 0.42 \,(\text{stat})\,  
       \pm 0.05 \,(\text{syst})\,
       ^{+0.09}_{-0.07} \,(\tau_{\Bc})\right) \times 10^{-3}}
}

\newcommand{\FinalResultEightTeV}{
  \ensuremath{\left(2.92  \pm 0.40 \,(\text{stat})\,  
       \pm 0.12 \,(\text{syst})\,
       ^{+0.21}_{-0.16} \,(\tau_{\Bc})\right) \times 10^{-3}}
}

\begin{document}

\renewcommand{\thefootnote}{\fnsymbol{footnote}}
\setcounter{footnote}{1}



\begin{titlepage}
\pagenumbering{roman}

\vspace*{-1.5cm}
\centerline{\large EUROPEAN ORGANIZATION FOR NUCLEAR RESEARCH (CERN)}
\vspace*{1.5cm}
\hspace*{-0.5cm}
\begin{tabular*}{\linewidth}{lc@{\extracolsep{\fill}}r}
\ifthenelse{\boolean{pdflatex}}
{\vspace*{-2.7cm}\mbox{\!\!\!\includegraphics[width=.14\textwidth]{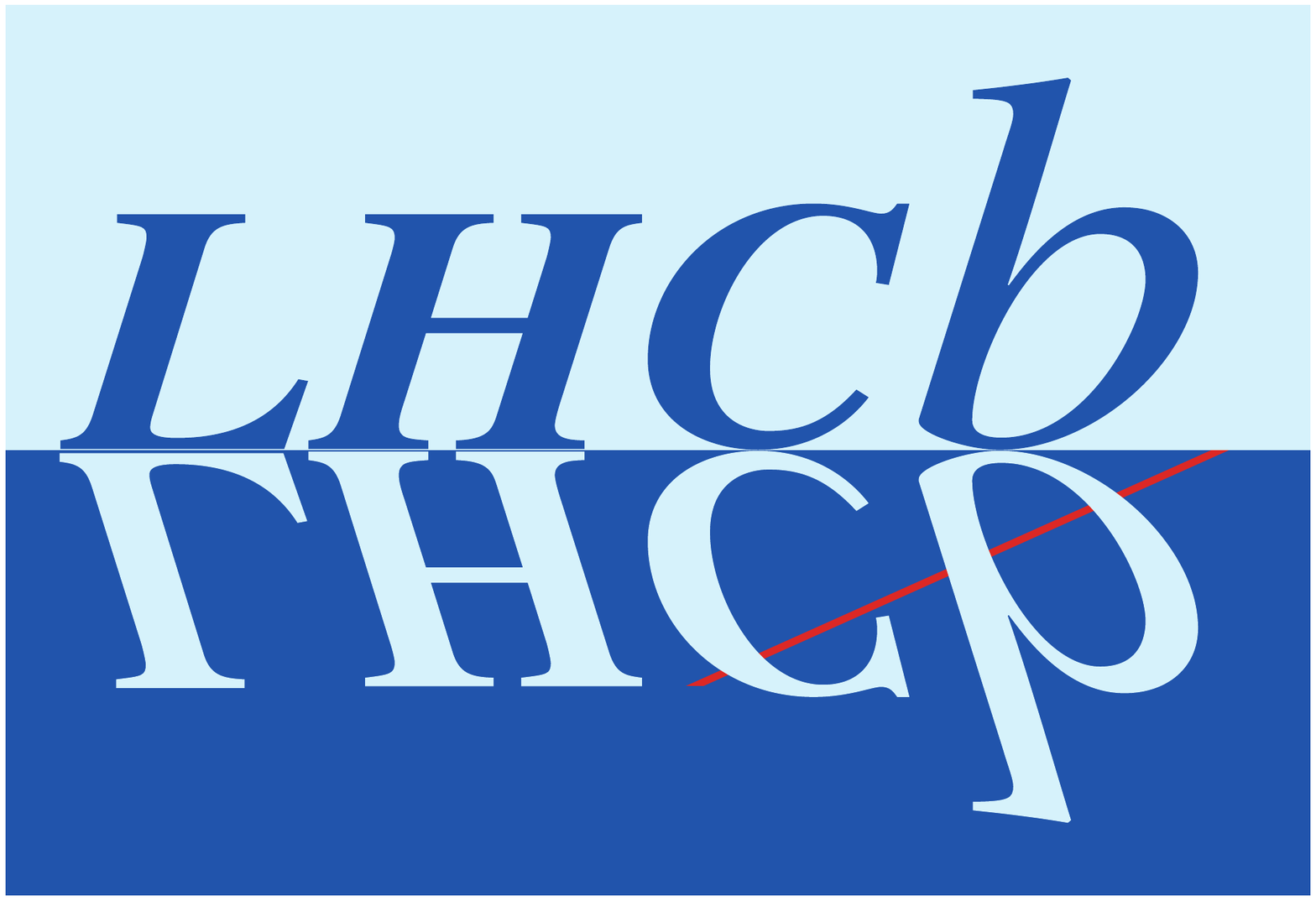}} & &}%
{\vspace*{-1.2cm}\mbox{\!\!\!\includegraphics[width=.12\textwidth]{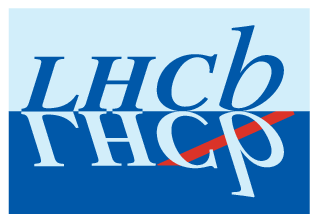}} & &}%
\\
 & & CERN-PH-EP-2013-136 \\  
 & & LHCb-PAPER-2013-044 \\  
 & & November 1, 2013
\end{tabular*}

\vspace*{2.0cm}

{\bf\boldmath\huge
\begin{center}
Observation of the decay \boldmath{$\Bcdec$}
\end{center}
}

\vspace*{2.0cm}

\begin{center}
The LHCb collaboration\footnote{Authors are listed on the following pages.}
\end{center}


\begin{abstract}
  \noindent
  The result of a search for the decay $\Bcdec$ is presented, using the $\Bsdecd$
  and $\Bsdecj$ channels. 
  The analysis is based on a data sample of 
  $pp$ collisions collected with the LHCb detector, corresponding to an integrated luminosity of
  $1\,\invfb$ taken at a center-of-mass energy of 7~TeV, and $2\,\invfb$ taken at 8~TeV.
  The decay $\Bcdec$ is observed with significance in excess of five standard deviations 
  independently in both decay channels. The measured product of the ratio of cross-sections and branching
  fraction is 
  $$
  \frat \times \mathcal{B}(\Bcdec) = \FinalResult ,
  $$  
  in the pseudorapidity range $2<\eta(B)<5$,
  where the first uncertainty is statistical, the second is systematic and the third is
  due to the uncertainty on the $\Bc$ lifetime. 
  This is the first observation of a $B$ meson
  decaying to another $B$ meson via the weak interaction.
\end{abstract}

\vspace*{2.0cm}

\begin{center}
  Published in Phys.~Rev.~Lett.~111,~181801~(2013)
\end{center}


{\footnotesize 
\centerline{\copyright~CERN on behalf of the \lhcb collaboration, license \href{http://creativecommons.org/licenses/by/3.0/}{CC-BY-3.0}.}}
\vspace*{2mm}

\end{titlepage}


\newpage
\setcounter{page}{2}
\mbox{~}
\newpage

\centerline{\large\bf LHCb collaboration}
\begin{flushleft}
\small
R.~Aaij$^{40}$, 
B.~Adeva$^{36}$, 
M.~Adinolfi$^{45}$, 
C.~Adrover$^{6}$, 
A.~Affolder$^{51}$, 
Z.~Ajaltouni$^{5}$, 
J.~Albrecht$^{9}$, 
F.~Alessio$^{37}$, 
M.~Alexander$^{50}$, 
S.~Ali$^{40}$
% ...

{\footnotesize \it
% ...
$ ^{5}$Clermont Universit\'{e}, Universit\'{e} Blaise Pascal, CNRS/IN2P3, LPC, Clermont-Ferrand, France\\
$ ^{6}$CPPM, Aix-Marseille Universit\'{e}, CNRS/IN2P3, Marseille, France\\
% ...
$ ^{9}$Fakult\"{a}t Physik, Technische Universit\"{a}t Dortmund, Dortmund, Germany\\
% ...
$ ^{36}$Universidad de Santiago de Compostela, Santiago de Compostela, Spain\\
$ ^{37}$European Organization for Nuclear Research (CERN), Geneva, Switzerland\\
% ...
$ ^{40}$Nikhef National Institute for Subatomic Physics, Amsterdam, The Netherlands\\
% ...
$ ^{45}$H.H. Wills Physics Laboratory, University of Bristol, Bristol, United Kingdom\\
% ...
$ ^{50}$School of Physics and Astronomy, University of Glasgow, Glasgow, United Kingdom\\
$ ^{51}$Oliver Lodge Laboratory, University of Liverpool, Liverpool, United Kingdom\\
% ...
}
\end{flushleft}

\cleardoublepage


\renewcommand{\thefootnote}{\arabic{footnote}}
\setcounter{footnote}{0}



\pagestyle{plain} 
\setcounter{page}{1}
\pagenumbering{arabic}

\begin{figure}[!b]
  \begin{center}
    \includegraphics[width=0.45\linewidth, bb= 125 569 256 677]{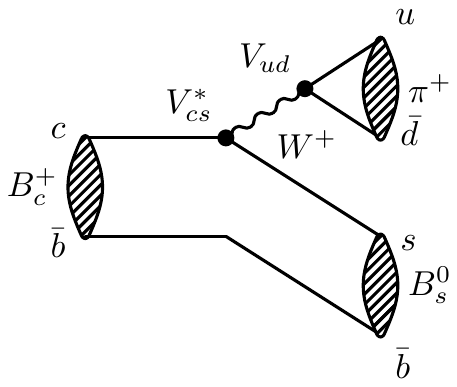}
    \vspace*{-0.5cm}
  \end{center}
  \caption{
    \small Leading-order Feynman diagram of the decay $\Bcdec$.}
  \label{fig:diagram}
\end{figure}

The $\Bc$ meson is the ground state of the $\bar{b}c$ system.
As such it is unique as it is the only weakly decaying doubly heavy meson.
All measurements of $\Bc$ meson decays to date are decays where the constituent $b$ quark decays weakly to a $c$ 
quark~\cite{Abe:1998wi, *Abe:1998fb, Aaltonen:2007gv, Abazov:2008kv, 
LHCb-PAPER-2012-028, LHCb-PAPER-2011-044, LHCb-PAPER-2012-054, LHCb-PAPER-2013-010, LHCb-PAPER-2013-021}. 
The decay of the $\Bc$ meson to another $B$ meson, with the bottom quark acting as a
spectator (see Fig.~\ref{fig:diagram}), has not previously been observed.
This will improve the understanding of theoretical predictions, 
and provide valuable information for the source of $\Bs$ mesons at the LHC.

A wide range of predictions for the branching fraction $\BR(\Bcdec)$
exists, between 16.4\,\% and 2.5\,\%, based on \eg QCD sum rules~\cite{Kiselev:2000pp,Gouz:2002kk},
or quark-potential models (see Refs.~\cite{Ivanov:2006ni,Anisimov:1998uk,Colangelo:1999zn,Ebert:2003wc,Dhir:2008zz,Naimuddin:2012dy}
and references therein). Experimental clarification is needed to shed light
on the present theoretical status.
Unlike most other $B$ decays, the higher order corrections in the expansion of Heavy Quark
Effective Theory within the framework of quantum chromodynamics (QCD) are relatively large. The
expansion is described in powers of $m_c/m_b$ rather than $\Lambda_{\rm QCD}/m_b$, due to the
presence of two heavy quark constituents, where $\Lambda_{\rm QCD}$ is the QCD scale, and $m_c$
($m_b$) the charm (bottom) quark mass.
In addition, the energy release in the decay is relatively small, leading to larger non-factorizable
effects compared to decays with lighter daughter particles.  Study of the decay
$\Bcdec$ allows these models to be tested.  Knowledge of the production of \Bs mesons
from $\Bc$ decays is also useful for time-dependent analyses 
of $\Bs$ decays, to understand any associated decay-time bias
due to the incorrect estimate of the $\Bs$ decay time if originating from a $\Bc$ decay, 
or to take advantage of flavor tagging capabilities using the accompanying
(``bachelor'') pion.

The data used in this analysis were collected with the LHCb detector~\cite{Alves:2008zz}
from $pp$ collisions 
at $\sqrt{s} = 7 \tev$ and $8 \tev$, corresponding to integrated luminosities of
1~$\invfb$ and 2~$\invfb$, respectively.  

The decays $\Bs \ra \Dsm \pip$ and $\Bs \ra \jpsi \phi$ are used, with the subsequent decays 
$\Dsdec$, $\jpsidec$ and $\phidec$. 
The inclusion of charge conjugate modes is implied throughout.
The event selection and fits to the $\Bs$ invariant mass distributions follow previous
LHCb analyses based on these $\Bs$ decay
modes~\cite{LHCb-PAPER-2012-037,LHCb-PAPER-2013-002}.
The two channels are analysed independently and the final results are combined.
The strategy is to normalize the final number of $\Bcdec$ decays to the number of 
$\Bs$ decays, which
gives a result for the $\Bcdec$ branching fraction multiplied by the ratio of 
$\Bc$ and $\Bs$ production rates, $(\sigma(\Bc)/\sigma(\Bs))\times \BR(\Bcdec)$.
The $\Bc$ signal region was not examined until the event selection was finalized.
Since the ratio of production rates, $\sigma(\Bc)/\sigma(\Bs)$, may depend on the kinematics
of the produced $B$ meson, the result is quoted for $B$ mesons produced in the pseudorapidity range
$2<\eta(B)<5$, corresponding to the LHCb detector acceptance.

The LHCb detector is a single-arm forward spectrometer covering the
pseudorapidity range $2<\eta<5$, described in detail in Ref.~\cite{Alves:2008zz}.
The combined tracking system provides momentum measurement with
relative uncertainty that varies from 0.4\,\% at $5 \gevc$ to 0.6\,\% at 
$100 \gevc$, and impact parameter
resolution of $20 \mum$ for tracks with high transverse
momentum, \pt. 
The impact parameter~(IP) is defined as the distance of closest 
approach between the track and a primary interaction.
Charged hadrons are identified using two ring-imaging Cherenkov
detectors. The charged pions from $\Bc$ decays are selected with efficiency of 93\,\%
while keeping the misidentification rate of kaons below 7\,\%. 
Muons are identified by a system composed of alternating layers of
iron and multiwire proportional chambers with a typical efficiency of 97\,\% at 1--3\,\%
pion to muon misidentification probability.  The trigger~\cite{LHCb-DP-2012-004}
consists of a hardware stage, based on information from the calorimeter and muon
systems, followed by a software stage, which applies a full event
reconstruction.
The $\Bs$ candidates with muons in the final state are required to pass the hardware trigger,
which selects muons with a transverse momentum, $\pt>1.48\gevc$,
whereas the $\Bs$ candidates with only hadrons in the final state 
are selected by requiring a hadron in the calorimeter with $\et>3.6\gevc$.

Monte Carlo simulations, used to develop the $\Bc$ candidate selection, are performed 
using \bcvegpy~\cite{Chang:2005hq}, interfaced with
\pythia~6.4~\cite{Sjostrand:2006za} using a specific \lhcb
configuration~\cite{LHCb-PROC-2010-056}.  Decays of hadronic particles
are described by \evtgen~\cite{Lange:2001uf}, 
in which final state radiation 
is generated using \photos~\cite{Golonka:2005pn}.
The interaction of the generated particles with the detector and its
response are implemented using the \geant
toolkit~\cite{Allison:2006ve, *Agostinelli:2002hh} as described in
Ref.~\cite{LHCb-PROC-2011-006}.

The $\Bs$ candidates are selected using the multi-variate analysis known as boosted
decision tree~(BDT)~\cite{Breiman, AdaBoost}, to optimally discriminate between
signal and background.  
In the training, simulated $\Bs$
decays are used as signal, whereas candidates in the $\Bs$ mass sideband
in data are used as background.  To avoid potential biases, only one sixth of
the data is used in the training.  It is verified that the distribution of the
BDT discriminant is the same for the events used in the training, compared to
those that were not.  All events are used for the final result.  The BDT
training for the selection of $\Bsdecd$ candidates uses only the upper sideband~$[5466, 5800]\,\mevcc$,
as the lower sideband contains a large amount of irreducible partially
reconstructed $B$ decays, while the training for $\BsToJPsiPhi$ uses both lower sideband $[5200, 5316]\,\mevcc$ 
and upper mass sideband ~$[5416, 5550]\,\mevcc$.  The $\Bs$ vertex quality ($\chi^2_{\rm vtx}$), flight
distance, momentum $p$ and $\pt$ are used to discriminate signal
from background. For the $\DsPi$ final state we use in addition the $\chi^2_{\rm
vtx}$, flight distance, $p$ and \pt of the $\Dsm$ candidate and the $p$, \pt and
$\chi^2_\textrm{IP}$ of the bachelor pion from the $\Bs$ decay to suppress
combinatorial background.  The quantity $\chi^2_\textrm{IP}$ is defined as the
difference in \chisq of a given primary vertex (PV) reconstructed with and
without the considered track.  The training for $\JpsiPhi$ candidates uses $p$,
\pt, $\chi^2_{\rm vtx}$ and $\chi^2_{\rm IP}$ of the $\jpsi$ and $\phi$
candidates, and the \pt of the final state kaons and muons.  In the selection of $\Bs$
candidates from $\Bc$ decays, variables that require the candidate to point to a
primary vertex, such as the impact parameter of the $\Bs$ candidate, are
explicitly not included.  The minimum value of the BDT discriminant is chosen by
optimizing the $\Bs$ signal significance $S/\sqrt{S+B}$, where $S$ and $B$ are
the expected numbers of signal and combinatorial background events,
respectively.

\begin{figure}[!t]
    \begin{picture}(500,150)(0,0)
     \put(0,1){\includegraphics[scale=0.50]{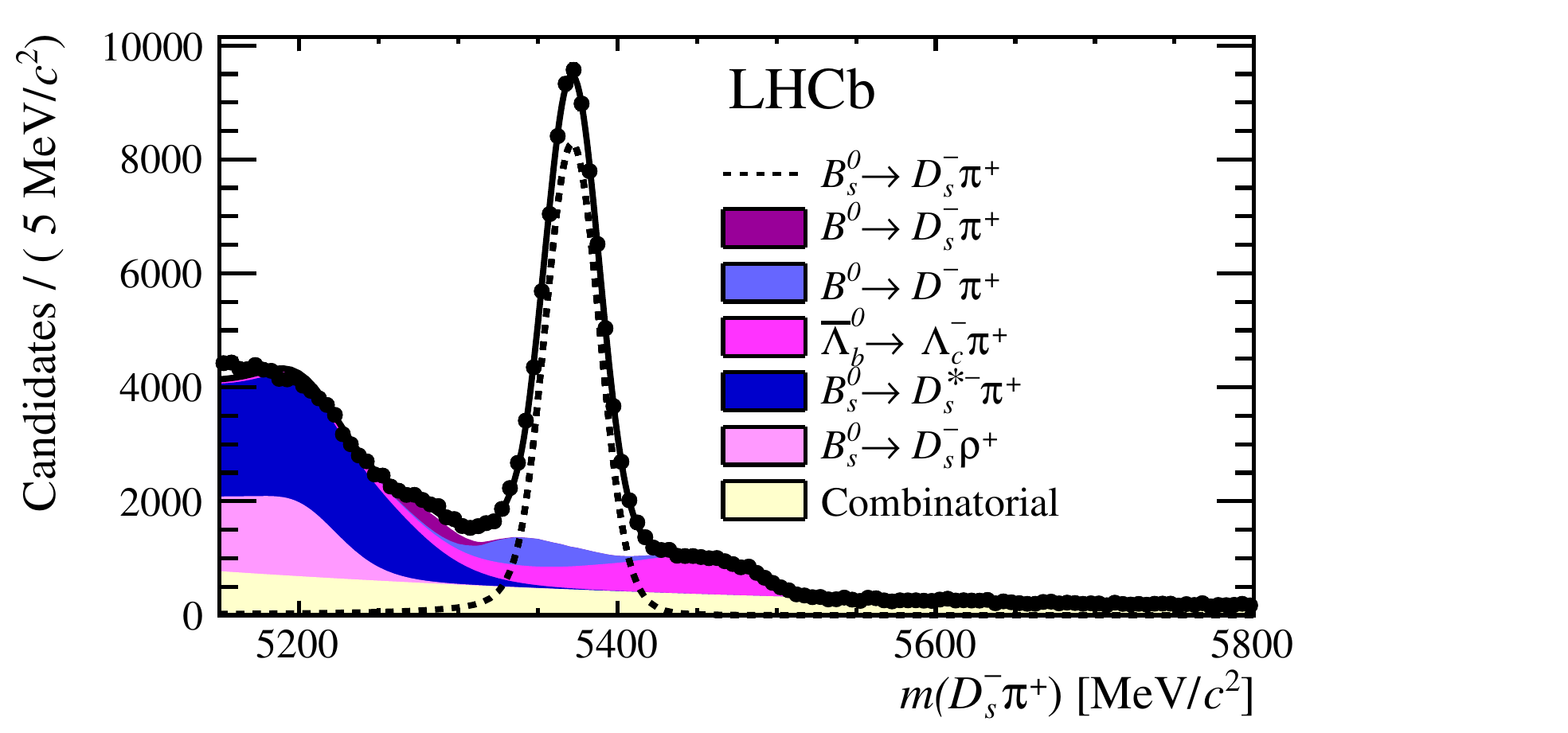}}
     \put(240,0){\includegraphics[scale=0.38]{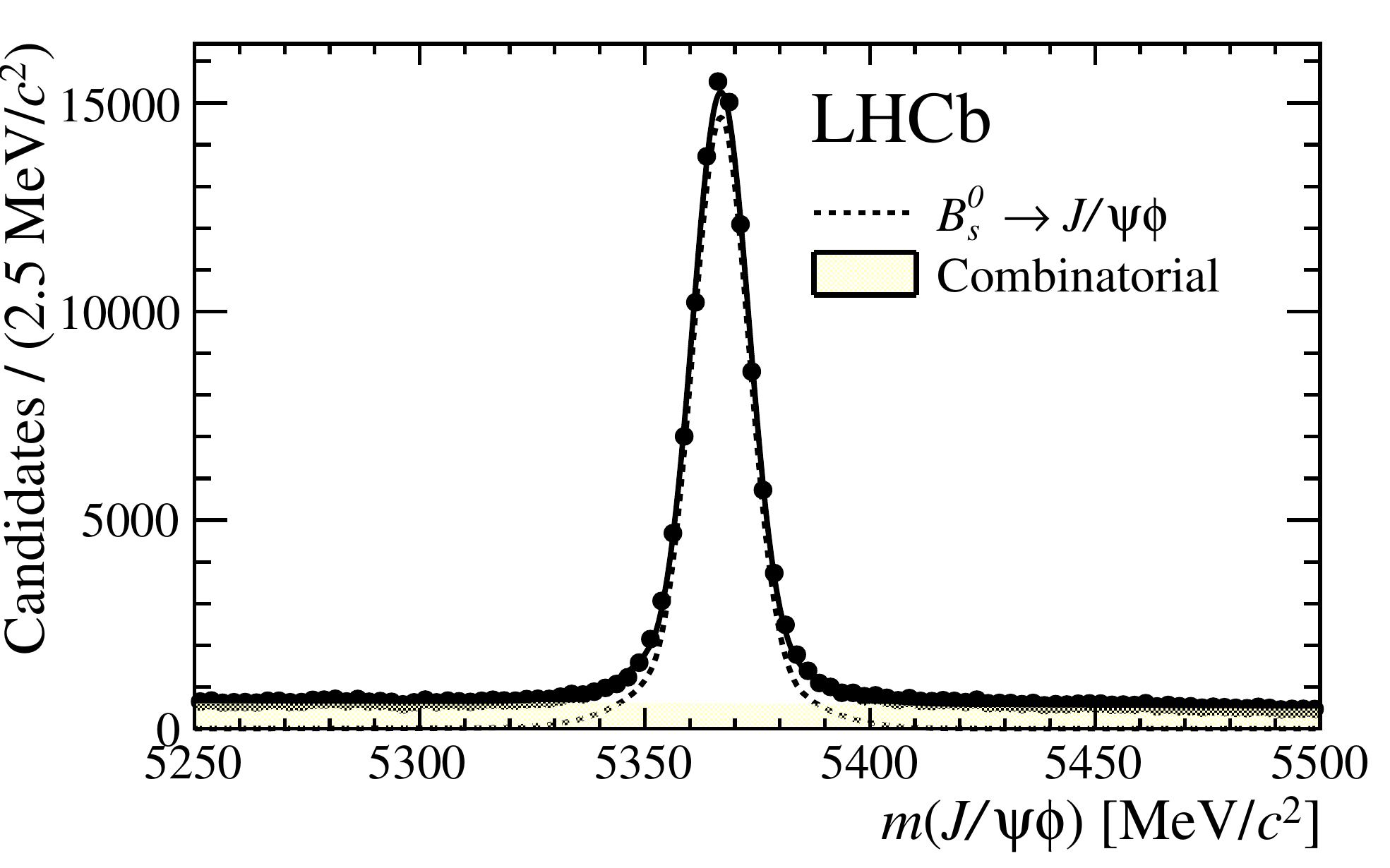}}
     \put(50,110){(a)}
    \put(290,110){(b)}
  \end{picture}
  \caption{\label{fig:BsMFit} Invariant mass distributions of 
    (a) $\Bsdecd$ and (b) $\Bsdecj$ candidates.
    The different components are defined in the legend.}
\end{figure}

The total number of $\Bs$ decays is obtained from extended unbinned maximum
likelihood fits to the invariant mass distributions, using mass constraints for
the $\jpsi$ candidates~\cite{Hulsbergen:2005pu}, and are shown in
Fig.~\ref{fig:BsMFit}. The signal shapes are taken as double Crystal
Ball functions~\cite{Skwarnicki:1986xj} with common peak value and with tails to either side of the peak, to
account for final state radiation and detector resolution effects. The
parameters that describe the tails are obtained from simulation and are fixed in the fits. 
The peak and width parameters of the signal are allowed to vary. The combinatorial
backgrounds are modeled with exponential distributions.  
The $\Bsdecd$ final state is contaminated by partially reconstructed $B$ decays such as $\Bs \ra \Dssm
\pip$ and $\Bs \ra \Dsm \rho^+$ decays, where the soft photon or neutral
pion is not reconstructed, and by decays where one of the final state particles is
misidentified as a kaon, such as $\Bz \ra \Dm\pip$ or $\bar{\PLambda}_b^0 \ra \PLambda_c^-\pip$ decays. 
The shapes of these backgrounds are fixed from
simulation, following Ref.~\cite{LHCb-PAPER-2012-037}.  In total
$103\,760 \pm 380$ $\Bsdecj$ and $73\,700\pm 500$ $\Bsdecd$ decays are found.

Selected $\Bs$ candidates with masses consistent with the known $\Bs$ mass are combined
with tracks that satisfy loose pion identification requirements.  Subsequently,
$\Bc$ candidates are selected with a second BDT algorithm.  
In the training of the second BDT, simulated candidates with masses consistent
with the $\Bc$ mass~\cite{PDG2012} are used as signal, and candidates in the
$\Bc$ mass sideband region in data are used as background.  For this, only the upper mass
sideband is used in the case of $\Bsdecd$, while also the lower mass
sideband is used in the case of $\Bsdecj$, to further suppress the larger combinatorial background
at smaller values of the mass.  Only one sixth of the total data set is used in the training.  The
second BDT uses the following variables: the $\Bc$ candidate
\pt, decay time, $\chi^2_{\rm vtx}$, $\chi^2_{\rm IP}$ and the $\Bc$ pointing
angle, \ie the angle between the $\Bc$ candidate momentum vector and the line
joining the associated PV and the $\Bc$ decay vertex.  The $\Bs$ polar angle (the
angle between $\Bs$ flight direction and the beam axis), decay time, decay
length and pointing angle are also used.  The $p$ and \pt of the bachelor pion
from the $\Bc$ decay are the most discriminating observables in the second BDT.
Differences between the analyses of the  $\DsPi$ and $\JpsiPhi$ final states are:
the use of $\chi^2_{\rm IP}$ of the $\Bs$ candidate and bachelor pion (from the $\Bc$ decay), 
and $\Bs$ and $\Bc$ momentum for the former;
the use of the $\Bc$ and $\Bs$ decay-length uncertainties for the latter.
The optimal selections are defined by maximizing figures of merit for a target
level of significance of three standard deviations, $\epsilon/(3/2 +
\sqrt{B})$~\cite{Punzi}, where $\epsilon$ is the signal efficiency for a given BDT criterion.
The figure of merit displays a plateau, and the chosen value is at the lower end
to allow to better constrain the shape of the combinatiorial background.  The
chosen selection is very close to the optimal point for a target level of
5\,$\sigma$ and for the expected significance $S/\sqrt{S+B}$. The trigger for $\Bsdecd$
decays preferentially selects candidates with high $\pt$ with respect to the
trigger for $\Bsdecj$ decays, which results in higher efficiency for the second
BDT requirement for the $\Bsdecd$ final state.  The $\Bc$ and $\Bs$ candidates
are required to be produced in the pseudorapidity range $2<\eta(B)<5$.

\begin{figure}[!b]
    \begin{picture}(500,150)(0,0)
     \put(-5,3){\includegraphics[scale=0.42]{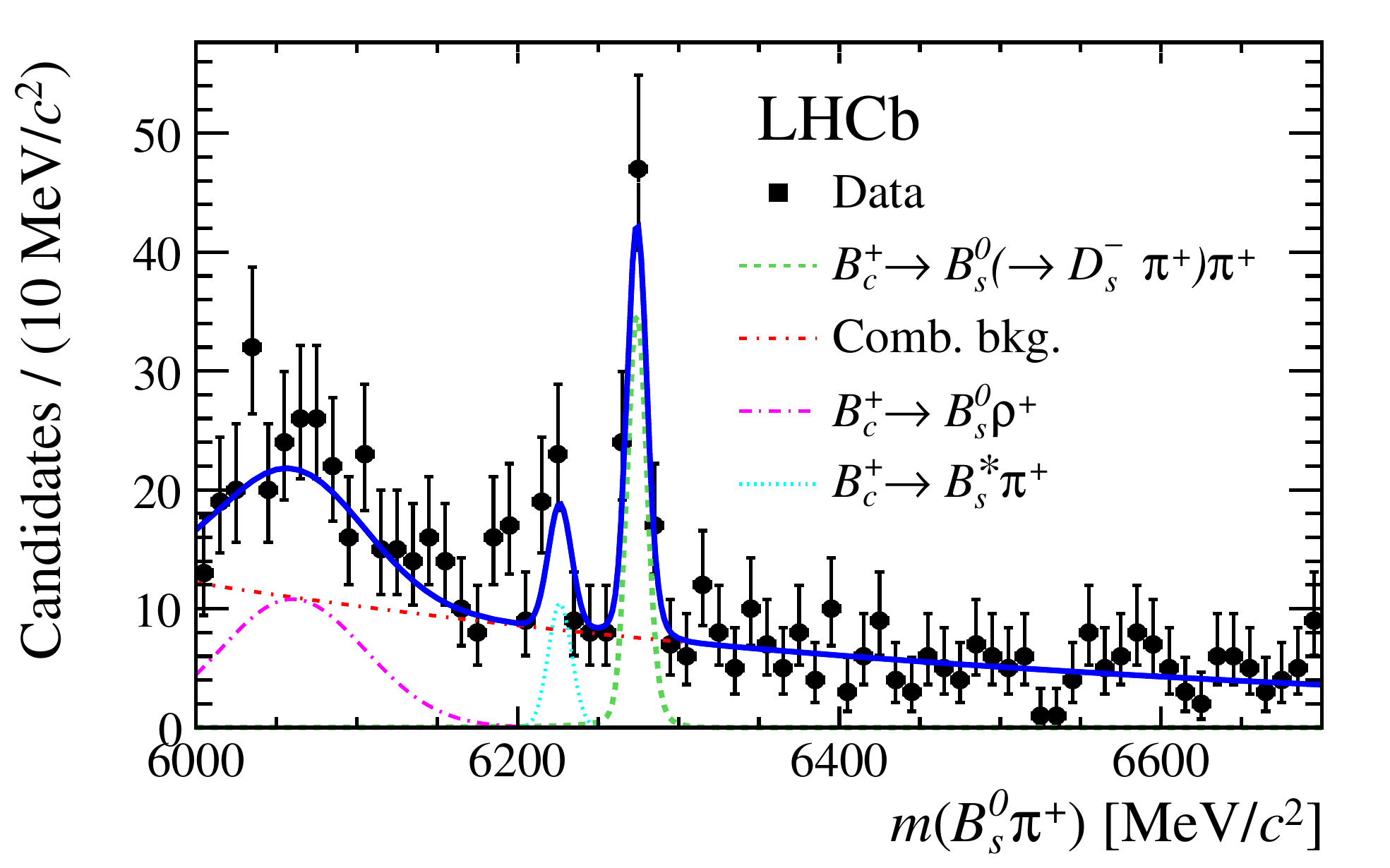}}
     \put(230,3){\includegraphics[scale=0.42]{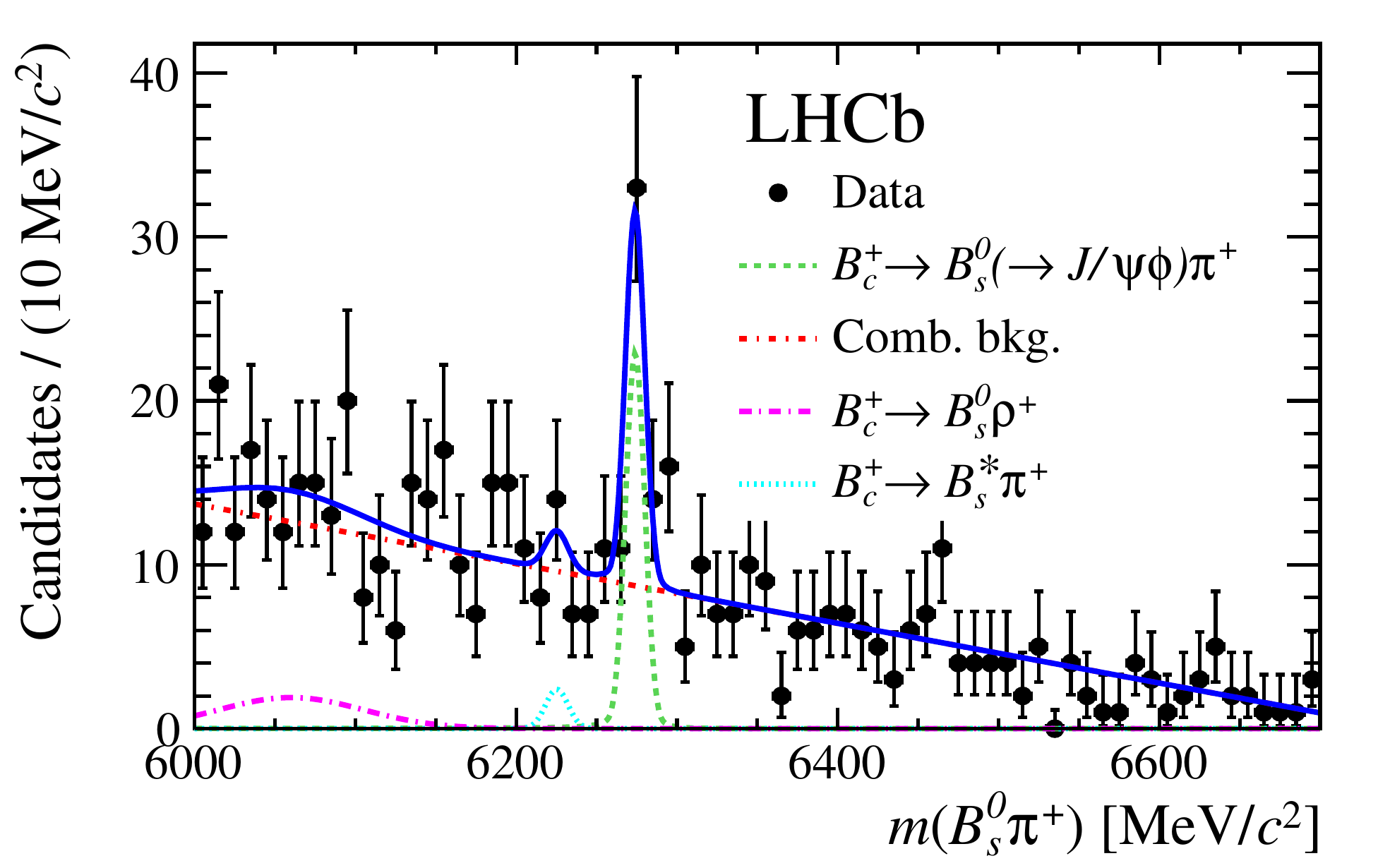}}
     \put(50,120){(a)}
    \put(290,120){(b)}
 \end{picture}
  \caption{\label{fig:BcMFit} $\Bc$ mass fits for the combined 2011 and 2012 data sets for 
(a) $\Bcdecfulld$ and (b) $\Bcdecfullj$ candidates.
The different components are indicated in the legends.}
\end{figure}

The invariant mass distributions for the $\Bcdec$ candidates are shown in
Fig.~\ref{fig:BcMFit}, together with the resulting fits.  The decay $\Bcdec$ has a
$Q$-value of 770\,\mevcc (with $Q\equiv m_{\Bc} - m_{\Bs} - m_{\pip}$), 
which results in a resolution of about 6~\mevcc when a
$\Bs$ mass constraint is applied.  The signal shape is modeled as a double
Crystal Ball function, with its parameters obtained from simulated events.  The
larger number of $\Bc$ candidates in the $\Bsdecd$ channel allows variation of
the peak position and the width in the fit.  The combinatorial background is
primarily due to signal $\Bs$ decays combined with a random pion from the
primary vertex, and is modeled with an exponential function. Backgrounds due to
$\Bc \ra \Bsstar \pip$ and $\Bc \ra \Bs \rho^+$ decays, where the photon or
neutral pion are not reconstructed, are simulated, and their shapes are modeled
with Gaussian distributions, with parameters fixed in the fit, and yields
allowed to vary.  Statistical signal significances of $7.7\,\sigma$ for
$\Bcdecfulld$ and $6.1\,\sigma$ for $\Bcdecfullj$ decays are obtained from the
likelihood ratio of fits with and without the probability density
function for the signal shape, 
$\sqrt{-2{\rm ln}(\mathcal{L}_{\rm B}/\mathcal{L}_{\rm S+B})}$,
with $64 \pm 10$ and $35 \pm 8$ signal decays,
respectively.

In Fig.~\ref{fig:BcMFit}a, 
the structure around $6225\mevcc$ is consistent with originating from $\Bc \ra \Bsstar \pip$ decays.
However, this contribution is not significant.

To obtain the value for the $\Bcdec$ branching fraction, multiplied by the
ratio of $\Bc$ and $\Bs$ production rates, the relative detection efficiency of
$\Bs$ decays compared to $\Bcdec$ decays is determined from simulation.
Requiring the bachelor pion to be inside the LHCb acceptance reduces the
$\Bcdec$ yield by about 19\,\% with respect to the $\Bs$ yield. The most significant
reduction in the number of selected $\Bc$ candidates comes from suppressing
$\Bs$ combinations with a random pion from the primary interaction, by means of
the second BDT selection.  The total relative detection efficiency of
$\Bcdec$ decays with respect to $\Bs$ decays is estimated to be 15.2\,\% for the
$\Bsdecj$ decay and 33.9\,\% for the $\Bsdecd$ final state.
This difference in $\Bc$ selection efficiencies is a consequence of the 
difference in $\Bs$ trigger and selection requirements.

\begin{table}[!b]
  \centering
  \caption{\label{tab:systlist} Contributions of the various sources of (relative) systematic uncertainty
on the efficiency-corrected ratio of event yields.
The total systematic uncertainty is the quadratic sum of the individual contributions.
The number of $\Bcdecfulld$ candidates is large enough that the peak position and width 
are freely varied in the fit, and hence the corresponding uncertainty is contained in the statistical
uncertainty of the signal yield.}
  \begin{tabular}{lcc}
\hline
Source                       & $\DsPi$ (\%)                     & $\JpsiPhi$ (\%) \\
\hline					       					 
$\Bs$ fit model              & 3.0                              & 1.2        \\
$\Bc$ mean mass              &   ...                            & 2.0        \\
$\Bc$ mass resolution        &   ...                            & 5.2        \\
$\Bc$ signal model           & 1.5                              & 1.7        \\
Combinatorial background model & 1.8                            & 0.3        \\
Partially reconstructed bkgd.& 1.8                              & 1.7        \\
Data-simulation difference   & 3.7                              & 3.7        \\ 
$\Bc$ lifetime               &\phantom{l}$^{+6.8}_{-3.5}$       & 7.4        \\ 
\hline					       					 
Total                        &\hspace{-0.25cm}\phantom{\Big(} $^{+8.9}_{-6.7}$  & 10.4       \\
\hline
  \end{tabular}
\end{table}

The sources of systematic uncertainty for the efficiency-corrected ratio
of $\Bc$ and $\Bs$ yields are listed in Table~\ref{tab:systlist}.
The uncertainty on the $\Bs$ yield in the $\DsPi$ analysis is determined by
varying the parameters that describe the tails of the signal mass distribution,
and by reducing the exponent of the combinatorial background by a factor two.
The uncertainty on the $\Bsdecj$ yield is obtained by comparing the
fitted yield in simulated pseudo-experiments to the yield that was used as input to
those experiments.

The uncertainty on the $\Bc$ yield is quantified by varying the peak position
and width in the fit to $\Bcdecfullj$ candidates. The signal model is validated
using simulated pseudo-experiments in the $\JpsiPhi$ analysis, whereas the tail parameters
are varied by $\pm 10\,\%$ in the $\DsPi$ analysis.  
In addition, the combinatorial background
shape is changed to a straight line, and the difference in the signal yield is
taken as the associated systematic uncertainty. The effect of partially
reconstructed $\Bc \ra \Bs \rho^+$ decays is estimated by excluding candidates
with mass less than $6150 \mevcc$ from the fit.  The significance of the
$\Bcdec$ signal is reduced to $7.5\,\sigma$ for $\Bcdecfulld$ and 
$5.5\,\sigma$ for $\Bcdecfullj$ when the systematic uncertainties on the fit to the $\Bc$ mass
distribution are taken into account.

The relative detection efficiency of $\Bc$ and $\Bs$ events is determined from
simulated events. The correspondence between data and simulation is quantified by
varying the criterion on the BDT value, and by comparing the observed $\Bs$ yield to the 
expected yield based on the change in efficiency as determined from simulation.
The largest contribution is due to the 10\,\% uncertainty on the \Bc lifetime~\cite{PDG2012},
which was recently improved by the CDF collaboration~\cite{Aaltonen:2012yb}.
The change in selection efficiency when varying the $\Bc$ lifetime by $\pm 10\,\%$
is assigned as systematic uncertainty.
A longer (shorter) $\Bc$ lifetime corresponds to a larger (smaller)
efficiency and therefore a smaller (larger) ratio.
As a cross-check, the effect of the choice of different sets of BDT input variables 
is investigated and the result is found to be stable.

The contribution from Cabibbo suppressed $\Bc \ra \Bs \Kp$ decays, 
the uncertainty on the efficiency of reconstructing the extra pion, and the uncertainty on the
efficiency of the particle identification requirement on the bachelor pion all give small
contributions ($<1.0\,\%$) to the total systematic uncertainty, 
and are not itemized in the summary in Table~\ref{tab:systlist}.

The $\Bs$ and $\Bc$ yields are corrected for the relative detection efficiencies, to obtain
the efficiency-corrected ratios of $\Bcdec$ over $\Bs$ yields,
$\BcBsPiDsPiResult$ and $\BcBsPiJpsiPhiResult$ for the $\DsPi$ and $\JpsiPhi$ final states,
respectively. 
The small fraction of $\Bs$ candidates originating from $\Bc$ decays is neglected.
The uncertainty due to the uncertainty on the $\Bc$ lifetime is correlated
between the two measurements, and is accounted for in the combined result
of the ratio of production rates multiplied with the branching fraction

$$
  \frat \times \mathcal{B}(\Bcdec) = \FinalResult,
$$
where the first uncertainty is statistical, the second is systematic and the third is
due to the uncertainty on the $\Bc$ lifetime.
Since $\sigma(\Bc)/\sigma(\Bs)$ may depend on the kinematics of the produced $B$ meson, 
the data are divided according to center-of-mass energy leading to \FinalResultSevenTeV
and \FinalResultEightTeV for $\sqrt{s} = 7$ and 8 TeV $pp$ collisions, respectively. 
The lower value for the result of the 7~TeV data
is attributed to a downward statistical fluctuation of the $\Bcdecfullj$ yield
in the 2011 data set, with a p-value of 1.5\,\%.

Assuming a value for $\BR(\Bc\ra \jpsi\pip)$ around 0.15\,\%~\cite{Ivanov:2006ni},
combined with the results 
$(\sigma(\Bc)/\sigma(\Bu))\times \BR(\Bc\ra \jpsi \pip)/\BR(\Bu\ra \jpsi \Kp) = 
(0.68 \pm 0.10 \pm 0.03 \pm 0.05)\,\%$~\cite{LHCb-PAPER-2012-028},
and measurements of $f_s/f_d$~\cite{LHCb-PAPER-2012-037} and $\BR(\Bu\ra \jpsi\Kp)$~\cite{PDG2012},
results in a ratio of production rates of $\Bc$ mesons over $\Bs$ mesons of about 0.02.
This leads to a branching fraction for $\Bcdec$ of about 10\,\%.  
Although precise quantification requires improved understanding of $\sigma(\Bc)$ and 
$\BR(\Bc\to \jpsi \pi^+)$,
even taking the lower estimates for $\BR(\Bc\to \jpsi \pi^+)$ that are found in the
literature~\cite{Ivanov:2006ni}, leads to a value of
$\BR(\Bcdec)$ which is the largest exclusive branching fraction of any known
weak $B$ meson decay.  

In summary, the first observation of a weak decay of a $B$ meson to another $B$
meson is reported.  This measurement will help to better understand flavor
tagging and the decay time resolution in time-dependent $\Bs$ analyses, and in
addition will constrain models that predict branching fractions of $\Bc$
decays.

\section*{Acknowledgements}
 
\noindent 
We wish to thank A.K.~Likhoded for useful discussions.
We express our gratitude to our colleagues in the CERN
accelerator departments for the excellent performance of the LHC. We
thank the technical and administrative staff at the LHCb
institutes. We acknowledge support from CERN and from the national
agencies: CAPES, CNPq, FAPERJ and FINEP (Brazil); NSFC (China);
CNRS/IN2P3 and Region Auvergne (France); BMBF, DFG, HGF and MPG
(Germany); SFI (Ireland); INFN (Italy); FOM and NWO (The Netherlands);
SCSR (Poland); MEN/IFA (Romania); MinES, Rosatom, RFBR and NRC
``Kurchatov Institute'' (Russia); MinECo, XuntaGal and GENCAT (Spain);
SNSF and SER (Switzerland); NAS Ukraine (Ukraine); STFC (United
Kingdom); NSF (USA). We also acknowledge the support received from the
ERC under FP7. The Tier1 computing centres are supported by IN2P3
(France), KIT and BMBF (Germany), INFN (Italy), NWO and SURF (The
Netherlands), PIC (Spain), GridPP (United Kingdom). We are thankful
for the computing resources put at our disposal by Yandex LLC
(Russia), as well as to the communities behind the multiple open
source software packages that we depend on.

\addcontentsline{toc}{section}{References}
\setboolean{inbibliography}{true}
\bibliographystyle{LHCb}
\bibliography{bc,main,LHCb-PAPER,LHCb-CONF,LHCb-DP}

\ifx\mcitethebibliography\mciteundefinedmacro
\PackageError{LHCb.bst}{mciteplus.sty has not been loaded}
{This bibstyle requires the use of the mciteplus package.}\fi
\providecommand{\href}[2]{#2}
\begin{mcitethebibliography}{10}
\mciteSetBstSublistMode{n}
\mciteSetBstMaxWidthForm{subitem}{\alph{mcitesubitemcount})}
\mciteSetBstSublistLabelBeginEnd{\mcitemaxwidthsubitemform\space}
{\relax}{\relax}

\bibitem{Abe:1998wi}
CDF collaboration, F.~Abe {\em et~al.},
  \ifthenelse{\boolean{articletitles}}{{\it {Observation of the $B_c$ meson in
  $p\bar{p}$ collisions at $\sqrt{s} = 1.8$ TeV}},
  }{}\href{http://dx.doi.org/10.1103/PhysRevLett.81.2432}{Phys.\ Rev.\ Lett.\
  {\bf 81} (1998) 2432}, \href{http://arxiv.org/abs/hep-ex/9805034}{{\tt
  arXiv:hep-ex/9805034}}\relax
\mciteBstWouldAddEndPuncttrue
\mciteSetBstMidEndSepPunct{\mcitedefaultmidpunct}
{\mcitedefaultendpunct}{\mcitedefaultseppunct}\relax
\EndOfBibitem
\bibitem{Abe:1998fb}
CDF collaboration, F.~Abe {\em et~al.},
  \ifthenelse{\boolean{articletitles}}{{\it {Observation of $B_c$ mesons in
  $p\bar{p}$ collisions at $\sqrt{s} = 1.8$ TeV}},
  }{}\href{http://dx.doi.org/10.1103/PhysRevD.58.112004}{Phys.\ Rev.\  {\bf
  D58} (1998) 112004}, \href{http://arxiv.org/abs/hep-ex/9804014}{{\tt
  arXiv:hep-ex/9804014}}\relax
\mciteBstWouldAddEndPuncttrue
\mciteSetBstMidEndSepPunct{\mcitedefaultmidpunct}
{\mcitedefaultendpunct}{\mcitedefaultseppunct}\relax
\EndOfBibitem
\bibitem{Aaltonen:2007gv}
CDF collaboration, T.~Aaltonen {\em et~al.},
  \ifthenelse{\boolean{articletitles}}{{\it {Observation of the decay
  $B_c^{\pm} \to J/\psi \pi^{\pm}$ and measurement of the $B_c^{\pm}$ mass}},
  }{}\href{http://dx.doi.org/10.1103/PhysRevLett.100.182002}{Phys.\ Rev.\
  Lett.\  {\bf 100} (2008) 182002}, \href{http://arxiv.org/abs/0712.1506}{{\tt
  arXiv:0712.1506}}\relax
\mciteBstWouldAddEndPuncttrue
\mciteSetBstMidEndSepPunct{\mcitedefaultmidpunct}
{\mcitedefaultendpunct}{\mcitedefaultseppunct}\relax
\EndOfBibitem
\bibitem{Abazov:2008kv}
D0 collaboration, V.~M. Abazov {\em et~al.},
  \ifthenelse{\boolean{articletitles}}{{\it {Observation of the $B_c$ meson in
  the exclusive decay $B_c$ $\to$ $J/\psi \pi$}},
  }{}\href{http://dx.doi.org/10.1103/PhysRevLett.101.012001}{Phys.\ Rev.\
  Lett.\  {\bf 101} (2008) 012001}, \href{http://arxiv.org/abs/0802.4258}{{\tt
  arXiv:0802.4258}}\relax
\mciteBstWouldAddEndPuncttrue
\mciteSetBstMidEndSepPunct{\mcitedefaultmidpunct}
{\mcitedefaultendpunct}{\mcitedefaultseppunct}\relax
\EndOfBibitem
\bibitem{LHCb-PAPER-2012-028}
LHCb collaboration, R.~Aaij {\em et~al.},
  \ifthenelse{\boolean{articletitles}}{{\it {Measurements of $B_c^+$ production
  and mass with the $B_c^+ \to J/\psi \pi^+$ decay}},
  }{}\href{http://dx.doi.org/10.1103/PhysRevLett.109.232001}{Phys.\ Rev.\
  Lett.\  {\bf 109} (2012) 232001}, \href{http://arxiv.org/abs/1209.5634}{{\tt
  arXiv:1209.5634}}\relax
\mciteBstWouldAddEndPuncttrue
\mciteSetBstMidEndSepPunct{\mcitedefaultmidpunct}
{\mcitedefaultendpunct}{\mcitedefaultseppunct}\relax
\EndOfBibitem
\bibitem{LHCb-PAPER-2011-044}
LHCb collaboration, R.~Aaij {\em et~al.},
  \ifthenelse{\boolean{articletitles}}{{\it {First observation of the decay
  $\Bc \to \jpsi \pi^+\pi^-\pi^+$}},
  }{}\href{http://dx.doi.org/10.1103/PhysRevLett.108.251802}{Phys.\ Rev.\
  Lett.\  {\bf 108} (2012) 251802}, \href{http://arxiv.org/abs/1204.0079}{{\tt
  arXiv:1204.0079}}\relax
\mciteBstWouldAddEndPuncttrue
\mciteSetBstMidEndSepPunct{\mcitedefaultmidpunct}
{\mcitedefaultendpunct}{\mcitedefaultseppunct}\relax
\EndOfBibitem
\bibitem{LHCb-PAPER-2012-054}
LHCb collaboration, R.~Aaij {\em et~al.},
  \ifthenelse{\boolean{articletitles}}{{\it {Observation of the decay $B^+_c
  \to \psi(2S)\pi^+$}},
  }{}\href{http://dx.doi.org/10.1103/PhysRevD.87.071103}{Phys.\ Rev.\  {\bf
  D87} (2013) 071103(R)}, \href{http://arxiv.org/abs/1303.1737}{{\tt
  arXiv:1303.1737}}\relax
\mciteBstWouldAddEndPuncttrue
\mciteSetBstMidEndSepPunct{\mcitedefaultmidpunct}
{\mcitedefaultendpunct}{\mcitedefaultseppunct}\relax
\EndOfBibitem
\bibitem{LHCb-PAPER-2013-010}
LHCb collaboration, R.~Aaij {\em et~al.},
  \ifthenelse{\boolean{articletitles}}{{\it {Observation of $B^+_c \rightarrow
  J/\psi D_s^+$ and $B^+_c \rightarrow J/\psi D_s^{*+}$ decays}},
  }{}\href{http://dx.doi.org/10.1103/PhysRevD.87.112012}{Phys.\ Rev.\  {\bf
  D87} (2013) 112012}, \href{http://arxiv.org/abs/1304.4530}{{\tt
  arXiv:1304.4530}}\relax
\mciteBstWouldAddEndPuncttrue
\mciteSetBstMidEndSepPunct{\mcitedefaultmidpunct}
{\mcitedefaultendpunct}{\mcitedefaultseppunct}\relax
\EndOfBibitem
\bibitem{LHCb-PAPER-2013-021}
LHCb collaboration, R.~Aaij {\em et~al.},
  \ifthenelse{\boolean{articletitles}}{{\it {First observation of the decay
  $B_c^+ \to J/\psi K^+$}},
  }{}\href{http://dx.doi.org/10.1007/JHEP09(2013)075}{JHEP {\bf 09} (2013)
  075}, \href{http://arxiv.org/abs/1306.6723}{{\tt arXiv:1306.6723}}\relax
\mciteBstWouldAddEndPuncttrue
\mciteSetBstMidEndSepPunct{\mcitedefaultmidpunct}
{\mcitedefaultendpunct}{\mcitedefaultseppunct}\relax
\EndOfBibitem
\bibitem{Kiselev:2000pp}
V.~Kiselev, A.~Kovalsky, and A.~Likhoded,
  \ifthenelse{\boolean{articletitles}}{{\it {$B_c$ decays and lifetime in QCD
  sum rules}}, }{}\href{http://dx.doi.org/10.1016/S0550-3213(00)00386-2}{Nucl.\
  Phys.\  {\bf B585} (2000) 353},
  \href{http://arxiv.org/abs/hep-ph/0002127}{{\tt arXiv:hep-ph/0002127}}\relax
\mciteBstWouldAddEndPuncttrue
\mciteSetBstMidEndSepPunct{\mcitedefaultmidpunct}
{\mcitedefaultendpunct}{\mcitedefaultseppunct}\relax
\EndOfBibitem
\bibitem{Gouz:2002kk}
I.~Gouz {\em et~al.}, \ifthenelse{\boolean{articletitles}}{{\it {Prospects for
  the $B_c$ studies at LHCb}},
  }{}\href{http://dx.doi.org/10.1134/1.1788046}{Phys.\ Atom.\ Nucl.\  {\bf 67}
  (2004) 1559}, \href{http://arxiv.org/abs/hep-ph/0211432}{{\tt
  arXiv:hep-ph/0211432}}\relax
\mciteBstWouldAddEndPuncttrue
\mciteSetBstMidEndSepPunct{\mcitedefaultmidpunct}
{\mcitedefaultendpunct}{\mcitedefaultseppunct}\relax
\EndOfBibitem
\bibitem{Ivanov:2006ni}
M.~A. Ivanov, J.~G. Korner, and P.~Santorelli,
  \ifthenelse{\boolean{articletitles}}{{\it {Exclusive semileptonic and
  nonleptonic decays of the $B_c$ meson}},
  }{}\href{http://dx.doi.org/10.1103/PhysRevD.73.054024}{Phys.\ Rev.\  {\bf
  D73} (2006) 054024}, \href{http://arxiv.org/abs/hep-ph/0602050}{{\tt
  arXiv:hep-ph/0602050}}\relax
\mciteBstWouldAddEndPuncttrue
\mciteSetBstMidEndSepPunct{\mcitedefaultmidpunct}
{\mcitedefaultendpunct}{\mcitedefaultseppunct}\relax
\EndOfBibitem
\bibitem{Anisimov:1998uk}
A.~Y. Anisimov, I.~Narodetsky, C.~Semay, and B.~Silvestre-Brac,
  \ifthenelse{\boolean{articletitles}}{{\it {The $B_c$ meson lifetime in the
  light front constituent quark model}},
  }{}\href{http://dx.doi.org/10.1016/S0370-2693(99)00273-7}{Phys.\ Lett.\  {\bf
  B452} (1999) 129}, \href{http://arxiv.org/abs/hep-ph/9812514}{{\tt
  arXiv:hep-ph/9812514}}\relax
\mciteBstWouldAddEndPuncttrue
\mciteSetBstMidEndSepPunct{\mcitedefaultmidpunct}
{\mcitedefaultendpunct}{\mcitedefaultseppunct}\relax
\EndOfBibitem
\bibitem{Colangelo:1999zn}
P.~Colangelo and F.~De~Fazio, \ifthenelse{\boolean{articletitles}}{{\it {Using
  heavy quark spin symmetry in semileptonic $B_c$ decays}},
  }{}\href{http://dx.doi.org/10.1103/PhysRevD.61.034012}{Phys.\ Rev.\  {\bf
  D61} (2000) 034012}, \href{http://arxiv.org/abs/hep-ph/9909423}{{\tt
  arXiv:hep-ph/9909423}}\relax
\mciteBstWouldAddEndPuncttrue
\mciteSetBstMidEndSepPunct{\mcitedefaultmidpunct}
{\mcitedefaultendpunct}{\mcitedefaultseppunct}\relax
\EndOfBibitem
\bibitem{Ebert:2003wc}
D.~Ebert, R.~Faustov, and V.~Galkin, \ifthenelse{\boolean{articletitles}}{{\it
  {Weak decays of the $B_c$ meson to $B_s$ and $B$ mesons in the relativistic
  quark model}}, }{}\href{http://dx.doi.org/10.1140/epjc/s2003-01347-5}{Eur.\
  Phys.\ J.\  {\bf C32} (2003) 29},
  \href{http://arxiv.org/abs/hep-ph/0308149}{{\tt arXiv:hep-ph/0308149}}\relax
\mciteBstWouldAddEndPuncttrue
\mciteSetBstMidEndSepPunct{\mcitedefaultmidpunct}
{\mcitedefaultendpunct}{\mcitedefaultseppunct}\relax
\EndOfBibitem
\bibitem{Dhir:2008zz}
R.~Dhir, N.~Sharma, and R.~Verma, \ifthenelse{\boolean{articletitles}}{{\it
  {Flavor dependence of $B_c$ meson form factors and $B_{c} \rightarrow PP$
  decays}}, }{}\href{http://dx.doi.org/10.1088/0954-3899/35/8/085002}{J.\
  Phys.\  {\bf G35} (2008) 085002}\relax
\mciteBstWouldAddEndPuncttrue
\mciteSetBstMidEndSepPunct{\mcitedefaultmidpunct}
{\mcitedefaultendpunct}{\mcitedefaultseppunct}\relax
\EndOfBibitem
\bibitem{Naimuddin:2012dy}
S.~Naimuddin {\em et~al.}, \ifthenelse{\boolean{articletitles}}{{\it
  {Nonleptonic two-body $B_c$-meson decays}},
  }{}\href{http://dx.doi.org/10.1103/PhysRevD.86.094028}{Phys.\ Rev.\  {\bf
  D86} (2012) 094028}\relax
\mciteBstWouldAddEndPuncttrue
\mciteSetBstMidEndSepPunct{\mcitedefaultmidpunct}
{\mcitedefaultendpunct}{\mcitedefaultseppunct}\relax
\EndOfBibitem
\bibitem{Alves:2008zz}
LHCb collaboration, A.~A. Alves~Jr. {\em et~al.},
  \ifthenelse{\boolean{articletitles}}{{\it {The \lhcb detector at the LHC}},
  }{}\href{http://dx.doi.org/10.1088/1748-0221/3/08/S08005}{JINST {\bf 3}
  (2008) S08005}\relax
\mciteBstWouldAddEndPuncttrue
\mciteSetBstMidEndSepPunct{\mcitedefaultmidpunct}
{\mcitedefaultendpunct}{\mcitedefaultseppunct}\relax
\EndOfBibitem
\bibitem{LHCb-PAPER-2012-037}
LHCb collaboration, R.~Aaij {\em et~al.},
  \ifthenelse{\boolean{articletitles}}{{\it {Measurement of the fragmentation
  fraction ratio $f_s/f_d$ and its dependence on $B$ meson kinematics}},
  }{}\href{http://dx.doi.org/10.1007/JHEP04(2013)001}{JHEP {\bf 04} (2013) 1},
  \href{http://arxiv.org/abs/1301.5286}{{\tt arXiv:1301.5286}}\relax
\mciteBstWouldAddEndPuncttrue
\mciteSetBstMidEndSepPunct{\mcitedefaultmidpunct}
{\mcitedefaultendpunct}{\mcitedefaultseppunct}\relax
\EndOfBibitem
\bibitem{LHCb-PAPER-2013-002}
LHCb collaboration, R.~Aaij {\em et~al.},
  \ifthenelse{\boolean{articletitles}}{{\it {Measurement of \CP-violation and
  the $B^0_s$-meson decay width difference with $B_s^0\to J/\psi K^+K^-$ and
  $B_s^0 \to J/\psi\pi^+\pi^-$ decays}},
  }{}\href{http://dx.doi.org/10.1103/PhysRevD.87.112010}{Phys.\ Rev.\  {\bf
  D87} (2013) 112010}, \href{http://arxiv.org/abs/1304.2600}{{\tt
  arXiv:1304.2600}}\relax
\mciteBstWouldAddEndPuncttrue
\mciteSetBstMidEndSepPunct{\mcitedefaultmidpunct}
{\mcitedefaultendpunct}{\mcitedefaultseppunct}\relax
\EndOfBibitem
\bibitem{LHCb-DP-2012-004}
R.~Aaij {\em et~al.}, \ifthenelse{\boolean{articletitles}}{{\it {The \lhcb
  trigger and its performance in 2011}},
  }{}\href{http://dx.doi.org/10.1088/1748-0221/8/04/P04022}{JINST {\bf 8}
  (2013) P04022}, \href{http://arxiv.org/abs/1211.3055}{{\tt
  arXiv:1211.3055}}\relax
\mciteBstWouldAddEndPuncttrue
\mciteSetBstMidEndSepPunct{\mcitedefaultmidpunct}
{\mcitedefaultendpunct}{\mcitedefaultseppunct}\relax
\EndOfBibitem
\bibitem{Chang:2005hq}
C.-H. Chang, J.-X. Wang, and X.-G. Wu,
  \ifthenelse{\boolean{articletitles}}{{\it {BCVEGPY2.0: An upgraded version of
  the generator BCVEGPY with the addition of hadroproduction of the $P$-wave
  $\Bc$ states}},
  }{}\href{http://dx.doi.org/10.1016/j.cpc.2005.09.008}{Comput.\ Phys.\
  Commun.\  {\bf 174} (2006) 241},
  \href{http://arxiv.org/abs/hep-ph/0504017}{{\tt arXiv:hep-ph/0504017}}\relax
\mciteBstWouldAddEndPuncttrue
\mciteSetBstMidEndSepPunct{\mcitedefaultmidpunct}
{\mcitedefaultendpunct}{\mcitedefaultseppunct}\relax
\EndOfBibitem
\bibitem{Sjostrand:2006za}
T.~Sj\"{o}strand, S.~Mrenna, and P.~Skands,
  \ifthenelse{\boolean{articletitles}}{{\it {PYTHIA 6.4 physics and manual}},
  }{}\href{http://dx.doi.org/10.1088/1126-6708/2006/05/026}{JHEP {\bf 05}
  (2006) 026}, \href{http://arxiv.org/abs/hep-ph/0603175}{{\tt
  arXiv:hep-ph/0603175}}\relax
\mciteBstWouldAddEndPuncttrue
\mciteSetBstMidEndSepPunct{\mcitedefaultmidpunct}
{\mcitedefaultendpunct}{\mcitedefaultseppunct}\relax
\EndOfBibitem
\bibitem{LHCb-PROC-2010-056}
I.~Belyaev {\em et~al.}, \ifthenelse{\boolean{articletitles}}{{\it {Handling of
  the generation of primary events in \gauss, the \lhcb simulation framework}},
  }{}\href{http://dx.doi.org/10.1109/NSSMIC.2010.5873949}{Nuclear Science
  Symposium Conference Record (NSS/MIC) {\bf IEEE} (2010) 1155}\relax
\mciteBstWouldAddEndPuncttrue
\mciteSetBstMidEndSepPunct{\mcitedefaultmidpunct}
{\mcitedefaultendpunct}{\mcitedefaultseppunct}\relax
\EndOfBibitem
\bibitem{Lange:2001uf}
D.~J. Lange, \ifthenelse{\boolean{articletitles}}{{\it {The EvtGen particle
  decay simulation package}},
  }{}\href{http://dx.doi.org/10.1016/S0168-9002(01)00089-4}{Nucl.\ Instrum.\
  Meth.\  {\bf A462} (2001) 152}\relax
\mciteBstWouldAddEndPuncttrue
\mciteSetBstMidEndSepPunct{\mcitedefaultmidpunct}
{\mcitedefaultendpunct}{\mcitedefaultseppunct}\relax
\EndOfBibitem
\bibitem{Golonka:2005pn}
P.~Golonka and Z.~Was, \ifthenelse{\boolean{articletitles}}{{\it {PHOTOS Monte
  Carlo: a precision tool for QED corrections in $Z$ and $W$ decays}},
  }{}\href{http://dx.doi.org/10.1140/epjc/s2005-02396-4}{Eur.\ Phys.\ J.\  {\bf
  C45} (2006) 97}, \href{http://arxiv.org/abs/hep-ph/0506026}{{\tt
  arXiv:hep-ph/0506026}}\relax
\mciteBstWouldAddEndPuncttrue
\mciteSetBstMidEndSepPunct{\mcitedefaultmidpunct}
{\mcitedefaultendpunct}{\mcitedefaultseppunct}\relax
\EndOfBibitem
\bibitem{Allison:2006ve}
Geant4 collaboration, J.~Allison {\em et~al.},
  \ifthenelse{\boolean{articletitles}}{{\it {Geant4 developments and
  applications}}, }{}\href{http://dx.doi.org/10.1109/TNS.2006.869826}{IEEE
  Trans.\ Nucl.\ Sci.\  {\bf 53} (2006) 270}\relax
\mciteBstWouldAddEndPuncttrue
\mciteSetBstMidEndSepPunct{\mcitedefaultmidpunct}
{\mcitedefaultendpunct}{\mcitedefaultseppunct}\relax
\EndOfBibitem
\bibitem{Agostinelli:2002hh}
Geant4 collaboration, S.~Agostinelli {\em et~al.},
  \ifthenelse{\boolean{articletitles}}{{\it {Geant4: a simulation toolkit}},
  }{}\href{http://dx.doi.org/10.1016/S0168-9002(03)01368-8}{Nucl.\ Instrum.\
  Meth.\  {\bf A506} (2003) 250}\relax
\mciteBstWouldAddEndPuncttrue
\mciteSetBstMidEndSepPunct{\mcitedefaultmidpunct}
{\mcitedefaultendpunct}{\mcitedefaultseppunct}\relax
\EndOfBibitem
\bibitem{LHCb-PROC-2011-006}
M.~Clemencic {\em et~al.}, \ifthenelse{\boolean{articletitles}}{{\it {The \lhcb
  simulation application, \gauss: design, evolution and experience}},
  }{}\href{http://dx.doi.org/10.1088/1742-6596/331/3/032023}{{J.\ Phys.\ Conf.\
  Ser.\ } {\bf 331} (2011) 032023}\relax
\mciteBstWouldAddEndPuncttrue
\mciteSetBstMidEndSepPunct{\mcitedefaultmidpunct}
{\mcitedefaultendpunct}{\mcitedefaultseppunct}\relax
\EndOfBibitem
\bibitem{Breiman}
L.~Breiman, J.~H. Friedman, R.~A. Olshen, and C.~J. Stone, {\em Classification
  and regression trees}, Wadsworth international group, Belmont, California,
  USA, 1984\relax
\mciteBstWouldAddEndPuncttrue
\mciteSetBstMidEndSepPunct{\mcitedefaultmidpunct}
{\mcitedefaultendpunct}{\mcitedefaultseppunct}\relax
\EndOfBibitem
\bibitem{AdaBoost}
R.~E. Schapire and Y.~Freund, \ifthenelse{\boolean{articletitles}}{{\it A
  decision-theoretic generalization of on-line learning and an application to
  boosting}, }{}\href{http://dx.doi.org/10.1006/jcss.1997.1504}{Jour.\ Comp.\
  and Syst.\ Sc.\  {\bf 55} (1997) 119}\relax
\mciteBstWouldAddEndPuncttrue
\mciteSetBstMidEndSepPunct{\mcitedefaultmidpunct}
{\mcitedefaultendpunct}{\mcitedefaultseppunct}\relax
\EndOfBibitem
\bibitem{Hulsbergen:2005pu}
W.~D. Hulsbergen, \ifthenelse{\boolean{articletitles}}{{\it {Decay chain
  fitting with a Kalman filter}},
  }{}\href{http://dx.doi.org/10.1016/j.nima.2005.06.078}{Nucl.\ Instrum.\
  Meth.\  {\bf A552} (2005) 566},
  \href{http://arxiv.org/abs/physics/0503191}{{\tt
  arXiv:physics/0503191}}\relax
\mciteBstWouldAddEndPuncttrue
\mciteSetBstMidEndSepPunct{\mcitedefaultmidpunct}
{\mcitedefaultendpunct}{\mcitedefaultseppunct}\relax
\EndOfBibitem
\bibitem{Skwarnicki:1986xj}
T.~Skwarnicki, {\em {A study of the radiative cascade transitions between the
  Upsilon-prime and Upsilon resonances}}, PhD thesis, Institute of Nuclear
  Physics, Krakow, 1986,
  {\href{http://inspirehep.net/record/230779/files/230779.pdf}{DESY-F31-86-02}}\relax
\mciteBstWouldAddEndPuncttrue
\mciteSetBstMidEndSepPunct{\mcitedefaultmidpunct}
{\mcitedefaultendpunct}{\mcitedefaultseppunct}\relax
\EndOfBibitem
\bibitem{PDG2012}
Particle Data Group, J.~Beringer {\em et~al.},
  \ifthenelse{\boolean{articletitles}}{{\it {\href{http://pdg.lbl.gov/}{Review
  of particle physics}}},
  }{}\href{http://dx.doi.org/10.1103/PhysRevD.86.010001}{Phys.\ Rev.\  {\bf
  D86} (2012) 010001}\relax
\mciteBstWouldAddEndPuncttrue
\mciteSetBstMidEndSepPunct{\mcitedefaultmidpunct}
{\mcitedefaultendpunct}{\mcitedefaultseppunct}\relax
\EndOfBibitem
\bibitem{Punzi}
G.~Punzi, \ifthenelse{\boolean{articletitles}}{{\it {Sensitivity of searches
  for new signals and its optimization}}, }{} in {\em Statistical Problems in
  Particle Physics, Astrophysics, and Cosmology} (L.~{Lyons}, R.~{Mount}, and
  R.~{Reitmeyer}, eds.), p.~79, 2003.
\newblock \href{http://arxiv.org/abs/physics/0308063}{{\tt
  arXiv:physics/0308063}}\relax
\mciteBstWouldAddEndPuncttrue
\mciteSetBstMidEndSepPunct{\mcitedefaultmidpunct}
{\mcitedefaultendpunct}{\mcitedefaultseppunct}\relax
\EndOfBibitem
\bibitem{Aaltonen:2012yb}
CDF collaboration, T.~Aaltonen {\em et~al.},
  \ifthenelse{\boolean{articletitles}}{{\it {Measurement of the $B_c^{-}$ meson
  lifetime in the decay $B_{c}^{-} \rightarrow J/\psi~\pi^{-}$}},
  }{}\href{http://dx.doi.org/10.1103/PhysRevD.87.011101}{Phys.\ Rev.\  {\bf
  D87} (2013) 011101}, \href{http://arxiv.org/abs/1210.2366}{{\tt
  arXiv:1210.2366}}\relax
\mciteBstWouldAddEndPuncttrue
\mciteSetBstMidEndSepPunct{\mcitedefaultmidpunct}
{\mcitedefaultendpunct}{\mcitedefaultseppunct}\relax
\EndOfBibitem
\end{mcitethebibliography}

\end{document}